\documentclass[sigconf]{acmart}

\acmConference[ICSE 2024]{46th International Conference on Software Engineering}{April 2024}{Lisbon, Portugal}

\renewcommand\footnotetextcopyrightpermission[1]{} % removes footnote with conference information in first column
\AtBeginDocument{%
  }

\usepackage{booktabs}
\usepackage{filecontents}

\usepackage{graphicx}
\usepackage[many]{tcolorbox}

\usepackage{mathtools,amssymb,latexsym,amsfonts,stmaryrd}
\usepackage{pbox}
\usepackage{xfrac}
\usepackage{booktabs}
\usepackage{xcolor}
\usepackage{threeparttable}
\usepackage{amsthm}
\usepackage{float}
\usepackage{multirow}
\usepackage{tikz}
\usepackage{mathrsfs}
\usepackage{mathpartir}
\usepackage{algorithm}
\usepackage{algorithmicx}
\usepackage{url}
\usepackage{syntax}
\usepackage{framed}
\usepackage[noend]{algpseudocode}
\usepackage{flushend}
\usepackage{listings}
\usepackage{moresize}
\usepackage{wrapfig}
 
\usepackage{seqsplit}
\usepackage{alltt}
\usepackage{xspace}
\usepackage{enumitem}

\newcommand{\revise}[2]{{\color{red}{\ifx&#1&\else- #1\fi}} {\color{ForestGreen}{\ifx&#2&\else+ #2\fi}}}%
\renewcommand{\revise}[2]{#2}%

\usetikzlibrary{arrows,patterns, decorations.pathreplacing}

\newcommand{\F}{Fig.}

\newcommand{\T}{Table}
\renewcommand{\S}{Sec.}

\newcommand{\ignore}[1]{}

\usepackage{amsmath, listings, amsthm, amssymb, proof, xspace}
\usepackage{bbding}

\newcommand{\CBrush}{\textcolor[RGB]{84,130,53}{\Checkmark}}
\newcommand{\XBrush}{\textcolor[RGB]{176,35,24}{\XSolidBrush}}
\newcommand{\TriUp}{\textcolor[RGB]{0,112,192}{$\Diamond$}}

\newcommand{\finding}[2]{
  \smallskip
  \smallskip
\begin{tcolorbox}[width=\linewidth,boxrule=0pt,top=1pt, bottom=1pt, left=1pt,right=1pt, colback=gray!20,colframe=gray!20]
\textbf{Finding #1:} %\textit
{#2}
\end{tcolorbox}}

\newcommand{\parh}[1]{\smallskip\noindent\textbf{#1}}
\newcommand{\sparh}[1]{\noindent\underline{#1}}
\begin{document}

%%
%% The "title" command has an optional parameter,
%% allowing the author to define a "short title" to be used in page headers.
\title{On Extracting Specialized Code Abilities from Large Language Models: A Feasibility Study}

\author{Zongjie Li}
\affiliation{%
 \institution{The Hong Kong University of Science and Technology}
 \city{Hong Kong SAR}
 \country{China}}
\email{zligo@cse.ust.hk}

\author{Chaozheng Wang}
\affiliation{%
 \institution{Harbin Institute of Technology}
 \city{Shenzhen}
 \country{China}}
\email{wangchaozheng@stu.hit.edu.cn}

\author{Pingchuan Ma}
\affiliation{%
 \institution{The Hong Kong University of Science and Technology}
 \city{Hong Kong SAR}
 \country{China}}
\email{pmaab@cse.ust.hk}

\author{Chaowei Liu}
\affiliation{%
 \institution{National University of Singapore}
 \city{Singapore}
 \country{Singapore}}
\email{e1011116@u.nus.edu}

\author{Shuai Wang}
\authornote{Corresponding authors.}
\affiliation{%
 \institution{The Hong Kong University of Science and Technology}
 \city{Hong Kong SAR}
 \country{China}}
\email{shuaiw@cse.ust.hk}

\author{Daoyuan Wu}
\affiliation{%
 \institution{Nanyang Technological University}
 \city{Singapore}
 \country{Singapore}}
\email{daoyuan.wu@ntu.edu.sg}
\authornotemark[1]

\author{Cuiyun Gao}
\affiliation{%
 \institution{Harbin Institute of Technology}
 \city{Shenzhen}
 \country{China}}
\email{gaocuiyun@hit.edu.cn}

% yangliu
\author{Yang Liu}
\affiliation{%
 \institution{Nanyang Technological University}
 \city{Singapore}
 \country{Singapore}}
\email{yangliu@ntu.edu.sg}

\begin{abstract}

    Recent advances in large language models (LLMs) significantly boost their usage in software engineering.
    However, 
    training a well-performing LLM demands a substantial workforce for data collection
    and annotation. 
    Moreover, training datasets may be proprietary or partially open,
    and the process often requires a costly GPU cluster.
    The intellectual property value of commercial LLMs makes them attractive targets for imitation attacks,
    but creating an imitation model with comparable parameters still incurs high costs.
    This motivates us to explore a practical and novel direction: \textit{slicing commercial black-box LLMs using medium-sized backbone models.}

    In this paper, we explore the feasibility of launching imitation attacks on LLMs to extract their \textit{specialized code abilities}, such as ``code synthesis'' and ``code translation.''
    We systematically investigate the effectiveness of launching code ability extraction
    attacks under different code-related tasks with multiple query schemes, including zero-shot, in-context, and Chain-of-Thought. We also
    design response checks to refine the outputs, leading to an effective
    imitation training process.
    Our results show promising outcomes, demonstrating that with a reasonable number of queries, attackers can train a medium-sized backbone model to replicate specialized code behaviors similar to the target LLMs.
    We summarize our findings and insights to help researchers better understand the threats posed by imitation attacks, including revealing a practical attack surface for generating adversarial code examples against LLMs.
\end{abstract}




% \received{20 February 2007}
% \received[revised]{12 March 2009}
% \received[accepted]{5 June 2009}

%%
%% This command processes the author and affiliation and title
%% information and builds the first part of the formatted document.
\maketitle

\section{Introduction}
\label{sec:introduction}

Recent advancements in the development of large language models (LLMs) have
led to a significant increase in their usage in software engineering~\cite{white2023chatgpt}. Major
enterprises, such as OpenAI~\cite{openai}, have already deployed their 
LLM APIs to assist humans in writing code and documents more
accurately and efficiently~\cite{copilot2022product}. A large-scale code corpus
with related natural language comments is used to build the models that
improve the productivity of computer programming. The advanced LLMs are
advocated to play a role as an ``AI programming assist'' that can handle
various code-related tasks. These tasks include both interactive
tasks, such as helping developers write automated
scripts~\cite{dalog2017deepcoder} or providing
reviews~\cite{codereview1,codereview2,codereview3}, and complex tasks that
require reasoning skills, such as finding vulnerabilities~\cite{devign2019zhou}
and clone detection~\cite{li2022unleashing}. Moreover, some LLMs like
ChatGPT~\cite{chatgpt} are designed with more interactive ability, which means
they can learn from their interactions with humans and improve their performance
over conversational turns, making them more effective at finding and fixing bugs that
are difficult to reach with traditional tools~\cite{xia2023conversational,ahmad2023fixing}.

Despite the growing popularity of LLM APIs, it is widely acknowledged that training a
well-performing LLM requires a plethora of workforce for data collection and annotation,
as well as massive GPU resources~\cite{brown2020language,chen2021evaluating}.
As a result, although some model architectures are publicly available, the model weights and training data
are viewed as intellectual property (IP) that model owners own and have rights over. 
The IP behind a model is quite valuable, as model providers offer their services to a large number of users.
Recently, it has been validated that computer vision (CV) and natural language processing (NLP)
models are vulnerable to \textit{imitation attacks} (or model extraction
attacks)~\cite{wallace2020imitation,xu2021student,xu2021beyond,he2022cater,DBLP:conf/aaai/HeXLWW22},
in which adversaries send carefully-designed queries to the victim
model and collect its responses to train a local model (often referred to as an
\textit{imitation model}) that mostly ``imitate'' the target remote model's
behavior.

At first glance, conducting an imitation attack on commercial LLMs may seem
impractical, as adversaries would need to prepare an imitation model with
comparable parameters, resulting in high costs. However, in contrast to the 
broad capabilities of LLMs, developers typically require only specialized 
subsets of these abilities, as they concentrate on particular tasks such as code translation and summarization.
This inspires us to explore a
practical and novel direction: slicing commercial, black-box LLMs using
medium-sized backbone models. In other words, we aim to
demonstrate the high possibility of extracting \textit{specialized code
abilities} of LLMs using \textit{medium-sized backbone models}. For instance,
attackers may be particularly interested in extracting the ability of ``code
translation'' from an LLM, which is a specialized code ability that allows the LLM
to translate source code from one programming language to another. This ability
is highly valuable in the software development industry and has been widely used
in commercial products~\cite{zheng2023codegeex,roziere2020unsupervised}.
The imitation attack also has the benefit of allowing users to avoid sharing their code snippets with third-party providers.
This is possible by locally deploying the specialized imitation model extracted from LLMs via medium-sized backbone models.

Extracting specialized code abilities poses unique technical challenges.
Depending on the specific code tasks,
LLMs may process natural language (NL) or programming language (PL) inputs and
emit NL/PL outputs accordingly. Moreover, modern LLMs can often be elicited by
various in-context~\cite{wei2022emergent} or Chain-of-Thought prompts~\cite{wei2022chain}, making it 
difficult to determine the attack surface of LLMs for code abilities. Furthermore, it is unclear how
to use the target LLM outputs to improve the performance of an imitation model,
given the complexity of code-related tasks and potential ambiguity in the outputs.

Overall, this work is the first to conduct a systematic and practical study on
the effectiveness of extracting specialized code abilities from LLMs using common
medium-sized models. To address the above technical challenges, we design and
conduct comprehensive imitation attacks on LLMs, including all three
code-related tasks and different query schemes (\S~\ref{subsec:query}).
Additionally, we develop several methods to refine the received outputs
of LLMs (\S~\ref{subsec:response}), which make the polished outputs more
effective in training the imitation model using two popular backbone models,
CodeT5~\cite{wang2021codet5} and CodeBERT~\cite{feng2020codebert}
(\S~\ref{subsec:modeltraining}). Finally, we demonstrate that the
imitation model can boost critical downstream tasks, e.g., adversarial example
generation (\S~\ref{subsec:appr-aegener}).

Based on the experimental results, we show that imitation attacks are effective for
LLMs on code-related tasks, and the performance of the imitation models surpasses that of the target LLMs
(with an average improvement of $10.33\%$). 
Moreover, we find substantial variance (from $4.94\%$ to $62.83\%$) in the impact of different query schemes across tasks. Queries providing adequate context generally improve performance on all tasks, while the Chain-of-Thought scheme only benefits code synthesis noticeably. These findings underscore the importance of selecting an appropriate query scheme.

We also explore the influence of different hyperparameters. 
We find that while $temperature$ and $top_p$ have an observable (yet not significant) impact,
the number of queries and in-context examples significantly affect the performance (from $41.97\%$ to $115.43\%$).
We recommend using three in-context examples for code-related tasks to balance the effectiveness and cost. 
Additionally, we demonstrate that the imitation model can assist in generating adversarial examples,
discovering up to 9.5\% more adversarial examples than the existing state-of-the-art models~\cite{wang2023robustness,jha2022codeattack,li2022cctest}.
Finally, we show the generalizability of our method on different LLM APIs, with the imitation models trained on gpt-3.5-turbo achieving approximately $92.84\%$ of the performance of those trained on text-davinci-003.

In summary, our contributions are as follows:

\begin{itemize}[leftmargin=*,noitemsep,topsep=0pt]
    \item We have conducted the first systematic study on the effectiveness of imitation attacks on a code knowledge slice of LLMs.
    \item Our study includes three representative query schemes
    and different code-related tasks. We have also designed several methods to refine the LLM
    outputs and enhance the training effectiveness of imitation models.
    \item We have formulated five research questions (RQs) to comprehensively evaluate
    the effectiveness of imitation attacks on LLMs for code abilities and their potential downstream applications,
    such as adversarial example generation.
    We have aggregated the results and observations to deduce empirical findings.
\end{itemize}

\section{Background and Related Work}
\label{sec:background}

\begin{table*}[!thp]
\caption{Benchmarking prior knowledge for imitation attacks. The symbols \CBrush, \TriUp,
      \XBrush\ denote require, partially require, and not require the corresponding knowledge when launching the model extraction, respectively.}
      \centering
\resizebox{0.8\linewidth}{!}{
\begin{tabular}{llllllll}
\hline
            & Data Distribution  & Model Architecture  & Output Probability                 & Object & Task & Victim model & Size \\ \hline
Chandrasekaran~\cite{chandrasekaran2020exploring}      & \CBrush        & \CBrush      & \CBrush  & -  & classification & - & -\\ \hline
Jagielski~\cite{jagielski2020high}      & \TriUp         & \CBrush      & \CBrush  & image   & classification & academic & 25.6M\\ \hline
Orekondy~\cite{orekondy2019knockoff}   & \TriUp        & \CBrush      & \CBrush  & image   & classification & academic & 21.8M\\ \hline
Yu et al.~\cite{yu2020cloudleak}     &  \TriUp       & \XBrush   & \XBrush  & image   & classification  &  commercial & 200M\\ \hline
He et al.~\cite{he2021model} &   \XBrush     & \XBrush     &  \TriUp & text   & generation  &  academic &  340M \\ \hline
Wallace~\cite{wallace2020imitation} &  \XBrush      & \XBrush     & \XBrush  & text   & generation  &  commercial &  -\\ \hline
Ours        &  \XBrush      & \XBrush     & \XBrush  & code   & generation  &  commercial &  175B\\ \hline
\end{tabular}
}
\label{tab:background}
\end{table*}

\subsection{Large Language Models for Code}
\label{subsec:inc-llm}

With a substantial amount of training resources (e.g., webtext
corpora~\cite{gao2020pile}) and model parameters (up to hundreds of
billions~\cite{brown2020language}), LLMs have manifested highly impressive performance on various tasks.
Typically, input training data is segmented into sentences and then into tokens, 
where each token consists of a sequence of characters before being fed to LLMs.
Previous works follow the classic ``pre-train and fine-tune'' paradigm~\cite{liu2021pre},
where a large number of datasets are used to train a general model as the public backbone, and users can fine-tune 
it with their private dataset and specialized task definition.
Following this paradigm, CodeBERT~\cite{feng2020codebert} and CodeT5~\cite{wang2021codet5} are two representative frameworks for code-related tasks. 

Moreover, for models scaled to 7B+ parameters\footnote{We refer to
the models with 7B+ parameters as large-sized models, 0.1B to 7B as medium-sized
models, and the rest are small-sized models.}, users can often guide them to
generate appropriate answers with texts referred to as \emph{prompts}. Thus, a
new paradigm called ``pre-train and prompt''~\cite{liu2021makes} has emerged.
This paradigm has been a huge success due to its powerful performance and high
flexibility. For example, OpenAI's ChatGPT~\cite{chatgpt} allows people to ask
it to complete different works with impressive accuracy and coherence given appropriate prompts. Due to
the drastic improvement caused by prompts, an increasing number of commercial
tools (e.g., Copilot~\cite{copilot}) are setting their backbones to LLMs that
follow the ``pre-train and prompt'' paradigm.

Note that not all LLMs possess capabilities for various code-related tasks, despite 
having substantial model parameters. The effectiveness of a
model is related to the datasets and methods used during training. For example,
the compile rate~\cite{chowdhery2022palm}, a popular metric assessing
the compilation correctness of LLCG outputs, for LaMDA's (137B) on DeepFIX is
only 4.3\%, whereas Codex (12B) gains 81.1\%. Therefore, this work only
considers LLMs that have demonstrated reasonable success on code-related tasks.

With the surge of Machine-Learning-as-a-Service (MLaaS), model owners often provide their services
via API or mature interface, and billing for queries can be broadly categorized as either monthly 
or ``pay-as-you-go.'' For example, GitHub Copilot charges 10 USD per month~\cite{CopilotBill}. In the latter charging
mode, users pay according to the number of queries they send or
the total length of tokens they receive from the API. 
For instance, j1-jumbo~\cite{AI21Bill} model from AI21 charges 0.03 USD per 1K tokens 
and 0.0003 USD per query.
All these models only share generated content with their users, and thus the owners can maintain the confidentiality of the 
underlying model architecture and training data.

\subsection{In-context Learning and Chain-of-Thought}
\label{subsec:inc-intro}

Mainstream LLMs significantly benefit from context prompts. 
It is shown that simply prompted with task definitions~\cite{liu2021pre} (also known as zero-shot learning) can often achieve 
satisfying results, and the performance can be further improved if more concrete in-context examples
exist~\cite{wei2022emergent}. Since the LLMs do not seem to share the same understanding of prompts with 
humans~\cite{DBLP:conf/naacl/WebsonP22,lu2021fantastically}, a series of works in prompt engineering have 
been proposed~\cite{wei2022chain, wang2022self, sanh2021multitask, DBLP:conf/emnlp/ShinRLWS20}.

Among them, one particular prompt strategy called Chain-of-Thought (CoT) 
has been shown to elicit strong reasoning abilities in LLMs by asking the model to incorporate intermediate 
reasoning steps (rationales) while solving a problem~\cite{wei2022chain,wang2022rationale,li2022explanations,li2022advance}.
Wang et al.~\cite{wang2022self} sample from the reasoning path and vote for the majority result.
Furthermore, Wang et al.~\cite{wang2022rationale} identify rationale sampling in the output space 
as the key component to robustly improve performance, and thus extending CoT to more tasks.
Such reasoning ability is not a feature only found in LLMs, and several works have explored
incorporating it in small models. Li et al.~\cite{li2022explanations} use CoT-like reasoning from LLMs to train 
smaller models on a joint task of generating the solution and explaining the solution generated. 
With multi-step reasoning, Fu et al.~\cite{fu2023specializing} concentrate small models on a specific task and sacrifice 
its generalizability in exchange for high performance.
To unleash the full potential of model extraction attacks, this work explores multiple
query schemes (including CoT) and quantifies their influence on performance.

\subsection{Imitation Attack}

\parh{Grey-Box Imitation Attack.}~The imitation attack, also known as model extraction attack, aims to emulate the
behavior of the victim model. Successfully extracting a model, especially
commercial APIs, is quite challenging, and previous
works~\cite{chandrasekaran2020exploring,jagielski2020high,yu2020cloudleak} tend
to simulate attacks in a grey-box setting, where different kinds of prior
knowledge are required. As shown in \T~\ref{tab:background}, this prior
knowledge includes the distribution of data, the model architecture, and the
output with its probability. For example, jagielski et
al.~\cite{jagielski2020high} require both prior knowledge like model
architecture and posterior knowledge such as the probability of output tokens to
aid in extraction. Moreover, since they mainly focus on the image
classification task, domain knowledge like class types can be used to boost
model extraction. For example, Yu et al.~\cite{yu2020cloudleak} ask for prior
knowledge of class types to establish a query set. 
He et al.~\cite{he2021model} demonstrate the vulnerability of powerful APIs built on fine-tuned BERT models to model extraction attacks,
which exploit the generated tokens and corresponding posterior probabilities.

\parh{Black-Box Imitation Attack.}~It is evident that prior knowledge of the target model is not always available in practice. LLM architecture (e.g., GPT-4~\cite{gpt4}) and training datasets are typically kept hidden by their owners.
With this regard, black-box imitation attack is more practical as it does not
require the adversaries to have any prior knowledge about the model internals
and its training data. 
To launch black-box imitation attack, attackers often first prepare a proxy
dataset, which is further derived into a query set $Q$ based on the API
documentation of the target LLM. Then, each query $q \in Q$ is sent to the
remote LLM API to obtain the corresponding output $o$. An imitation model can
be trained with the collected dataset $\{q_i, o_i | q_i \in Q, o_i \in O\}$.

\subsection{Adversarial Examples (AEs)}
\label{subsec:back-AE}

Adversarial attacks are known for their
ability to introduce imperceptible changes to input data, resulting in incorrect
output generated by a model. The inputs that induce failures are referred to as
AEs~\cite{goodfellow2014explaining}. 
While it is relatively easy to generate AEs for common deep neural networks, LLMs
are believed to be more resilient toward
AEs~\cite{wang2023robustness,bubeck2023sparks} due to the following reasons: (1)
LLM often exposes only APIs in a ``black-box'' setting, and (2) the model
vendors usually put a considerable amount of efforts into fine-tuning the model.
For instance, OpenAI has disclosed that it spent over six months making GPT-4
safer and more aligned, assembling a team of over a hundred domain experts
specializing in model alignment and adversarial testing before its public
release~\cite{oai-safety}.

Several works have been proposed to test and exploit the
potential of adversarial examples in models designed for code-related tasks. CodeAttack~\cite{jha2022codeattack}
collected logit information from CodeT5 and CodeBert, makes
small adjustments to code tokens, and generates AEs accordingly. Yang et
al.~\cite{yang2022important} proposed a search-based framework named Radar to
generate function names that cause AEs. CCTest~\cite{li2022cctest} mutated Python
code using several semantics-preserving transformations and detected inconsistent
outputs in Copilot over those mutated inputs. 

\begin{figure*}[!htpb]
      \centering
      \includegraphics[width=0.85\textwidth]{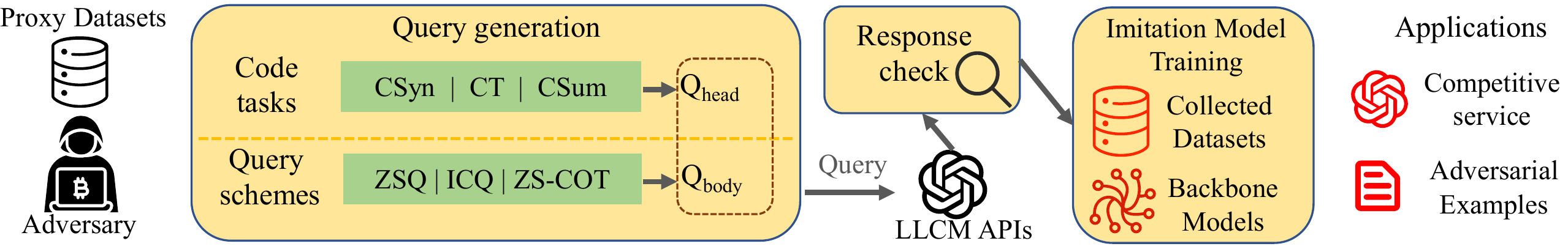}
      \vspace{-5pt}
      \caption{An overview of our imitation attack framework, including query generation, response check, imitation training, and downstream applications.}
      \label{fig:overview}
      \vspace{-5pt}
\end{figure*}

\section{Ethics and Responsible Disclosure}
\label{sec:ethics}

The aim of our work is to enhance the resilience of LLM APIs, and we strongly believe that providing democratized access to sophisticated LLM APIs benefits all of humanity.
To accomplish this goal, we conduct our experiments in a responsible manner and endeavor to minimize any potential real-world harm.

\parh{Minimal Real-world Harm.}\noindent~Since we discovered adversarial examples using our imitation model, we minimized harm by (1) avoiding any damage to
real users and (2) reporting potential adversarial examples to OpenAI, as well as the detailed attacking process and methods used to find them.

\parh{Responsible Data Disclosure.}\noindent~As in previous work~\cite{gpt4all,alpaca,vicuna2023}, 
we collected the dataset from OpenAI and used it to train the imitation model, which provides competitive 
services and aids in adversarial attacks. 
Therefore, to prevent potential misuse, we only make the scripts used for collecting the dataset publicly available and reveal the AEs that have already been fixed.
Furthermore, we adhere strictly to the licenses of the backbone models and proxy datasets, and no profit has been gained.

Overall, we believe that the security of LLM ecosystem is best advanced by responsible researchers and model owners surfacing these problems.

\section{Technical Pipeline}
\label{sec:approach}

In this work, we aim to launch imitation attacks to extract a "slicing" of code knowledge from LLMs using medium-sized backbone models. This task is challenging primarily due to LLMs can be queried in various ways, making it necessary to explore various query schemes. Their strong in-context understanding capabilities and flexibility across different tasks add to the complexity of benchmarking their attack surfaces. 
Specifically, LLMs have demonstrated strong in-context
understanding capabilities~\cite{liu2021pre}. With fewer examples provided,
LLMs can achieve a better understanding of downstream tasks with higher
performance~\cite{wei2022emergent}. Also, Chain-of-Thought
reasoning~\cite{wang2022self,wei2022chain} elicits the complex reasoning ability in LLMs. These novel tasks and schemes make LLMs quite flexible and versatile under various tasks, thus making it highly challenging and costly to systematically benchmark the attack surfaces of LLMs against model extraction.

\F~\ref{fig:overview} presents an overview of our imitation attack, which consists of four phases: (1) query
generation, (2) response checking, (3) imitation model training, and (4) downstream (adversarial) applications. Given one or more 
proxy datasets, our attack framework first generates LLM queries according to different code tasks and query schemes.
We then employ a rule-based filter to check
the correctness and quality of the responses provided by LLMs.
Responses that pass the filter are considered of high-quality and used to train the imitation model.
Next, we train the imitation model by fine-tuning medium-sized backbone models with the filtered responses.
Finally, we use the imitation model for various downstream (malicious) applications, such as providing competitive service and boosting the generation of AEs.
We now describe each phase in detail.

\subsection{Imitation Query Generation}
\label{subsec:query}

According to our preliminary tests, query schemes and prompt quality have a
significant impact on eliciting LLM outputs. Therefore, to fully unleash the
potential of this attack, we benchmark three query schemes as detailed below. Before introducing
them, we first clarify how a query is decomposed into two parts: the question
head $Q_{head}$ and the question body $Q_{body}$. Note that the former
varies from task to task, and the latter can be collected from
proxy datasets. For example, for the code summarization task, we set the sentence ``Summarize the following code in
one sentence'' as $Q_{head}$. This is beneficial
because a clear and concise question can help LLMs understand their specific role 
within a task, which in turn enhances their abilities to effectively complete that task.

\parh{Zero-Shot Query (ZSQ).}~This scheme specifies that attackers will
iteratively use each question $q_i \in Q$ to query the target LLM without
preparing any context. Accordingly, the attackers will gather the responses to
train the imitation model. ZSQ is universal and adopted by nearly all prior model
extraction works~\cite{orekondy2019knockoff,he2022cater,DBLP:conf/aaai/HeXLWW22,yu2020cloudleak}
in both classification and generation tasks. Since the tasks we evaluated (see
\T~\ref{tab:task}) are all generation tasks, we
follow~\cite{he2022cater,yu2020cloudleak} to prepare the queries.

\parh{In-context Query (ICQ).}~Several studies~\cite{lu2021fantastically,brown2020language,wei2022chain,lampinen2022can}
have shown that providing in-context information can significantly improve LLM 
performance. Therefore, we also consider this query scheme.
Specifically, for each question $q_i$, the query would provide several
examples along with the question head $Q_{head}$. The choice of examples used as
contexts plays a crucial role in enabling the model to understand the task.
For instance, the number of examples can not
be too few or too many because a short in-context may not provide enough
information to the victim model, while a long one may cause the query to exceed
the length limit.
We study its influence in \S~\ref{subsec:rq3}.

\parh{Zero-Shot CoT (ZS-COT).}~This scheme was first proposed
in~\cite{kojima2022large}. Unlike~\cite{wei2022chain}, which
heavily relies on manually crafted prompts, 
the core idea behind ZS-CoT is quite simple: adding a prompt sentence like
``\textit{Let's think it step by step.}'' to extract the reasoning process hidden
in the model. We follow the definition proposed by~\cite{kojima2022large} and
treat ZS-CoT as a two-step prompt process. Specifically, given a standard query sample $s_i \in
S$, which contains a question $q_i \in Q$, ZS-CoT first constructs a prompt that
requires explanation (or rationale) to the victim model and collects the
response as $r_i$. In the second stage, it uses both $q_i$ and $r_i$ to ask for
the final answer of victim models. Usually, the two-stage query sequence follows
this form: ``\textit{$Q_{head}$  Q: $\langle q_i \rangle$. A: Let's think it step
by step. $\langle r_{i}^{1} \rangle$. Therefore, the answer is $\langle
a_{i}^{2} \rangle$}.''\footnote{The superscript for $r_i$ and $a_i$ indicates
the stage at which they were collected. Readers can refer to~\cite{artifact} for details.} In the experiment, we slightly changed the
answer trigger pattern to adapt to the answer format. For example, we used
``\textit{Therefore, the translated C\# code is}'' for code translation task and
``\textit{Therefore, the summarization is}'' for code summarization task. 
It is worth noting that even with a carefully designed filtering system provided by~\cite{ho2022large},
it is hard to guarantee the correctness of the rationale. 
This is because the code-related tasks considered in this paper are open-ended, 
which poses a greater challenge than the multiple-choice questions in NLP reasoning tasks. 
As a result, we assess the generated rationales for each task (see \S~\ref{subsec:rq2} for details).
It is worth noting that we do not consider in-context CoT (IC-CoT) in this paper.
The main reason for this exclusion is the difficulty of constructing appropriate Chain-of-Thought reasoning processes~\cite{wei2022chain} for code-related tasks, as opposed to arithmetic or symbolic reasoning tasks.

\subsection{Response Check Scheme}
\label{subsec:response}

As mentioned earlier, the collected responses should be refined to achieve high effectiveness in training an imitation model. That said, when attackers receive
the response from LLMs, it is desirable to first perform a check before adding
it to the training data.
This can improve the overall quality of the dataset by helping attackers identify and rule out poor quality content, 
thereby increasing the average quality of responses. 
Additionally, trimming the training data can reduce the overall training cost of imitation models.
Specifically, given a list of LLM outputs $O^{L}$, we check each output $o^{L}_{i} \in
O^{L}$ based on several metrics and keep the high quality answers.

To handle LLMs that may produce both NL and PL outputs, we have devised distinct filtering rules for each. 
For NL outputs, we count the length and discard any text below or above pre-defined thresholds. Following the setting in
CodexGLUE~\cite{lu2021codexglue}, the upper bound is set to 256, and the lower bound is set to 3.

For PL outputs, we only keep those that pass a grammar check by a parser. For this purpose, we use
treesitter~\cite{treesitter}, a parser generator tool that can parse incomplete
code fragments. Unlike language-specific compilers or interpreters (e.g., CPython for Python),
treesitter supports most mainstream languages\footnote{Treesitter version 0.20.7
now supports 113 different programming languages.} with a unified interface,
which makes it convenient to adapt to various PLs.
Moreover, treesitter provides error messages for incorrect code syntax, which enables us to count the number of 
failures and further investigate the reasons behind them (see \S~\ref{subsec:rq2}).

\subsection{Imitation Model Training}
\label{subsec:modeltraining}

Similar to previous imitation attacks~\cite{wallace2020imitation,he2021model}, answers that pass the 
response selection module are considered as high-quality answers and used to train the imitation model.
Specifically, the collected response dataset $\{q_i,o_i|q_i\in Q, o_i \in O \}$ is used to fine-tune the public backbone model.
In this subsection, we first provide details on the target LLMs and their code-related tasks that are the focus of our imitation attacks. We then explain the evaluation metrics used and the process for training the imitation models.

\parh{Target LLMs.}~Unless otherwise specified, we use OpenAI's text-davinci-003~\cite{davinci}
as the victim LLM for all experiments, since 
it has been widely used in prior
research~\cite{bommarito2022gpt,wang2022towards}, which supports its
efficacy and reliability. In our evaluation (\S~\ref{subsec:rq5}), we further demonstrate the
generalizability of our attack by evaluating it with another LLM API called gpt-3.5-turbo.

\begin{table}[!thp]
  \caption{The evaluated tasks and datasets. CSyn, CT, and CSum denote code synthesis, code
  translation, and code summarization, respectively. CSN
  represents the CodeSearchNet dataset. $D_{proxy}$ and $D_{ref}$ are the proxy
  and reference datasets.}
  \label{tab:task}
  \centering
  \resizebox{0.9\linewidth}{!}{
  \begin{tabular}{lllll}
  \hline
  Category    & $D_{proxy}$                       & $D_{ref}$                          & \# Queries & Stat. of  $D_{ref}$     \\ \hline
  CSyn        & XLCOST~\cite{zhu2022xlcost}       & CONALA~\cite{yin2018mining}        & 2k & 2k/-/500                \\ 
  CT          & XLCOST~\cite{zhu2022xlcost}       & CodeXGLUE~\cite{lu2021codexglue}   & 10k & 10k/500/1k              \\
  CSum        & DualCODE~\cite{wei2019code}       & CSN~\cite{husain2019codesearchnet} & 8k & 25k/14k/15k             \\ \hline
  \end{tabular}
  }
  \end{table}

\parh{Target LLM Tasks.}~Before providing details on the filtering rules, we introduce the 
target LLM tasks in this research.
Drawing from the variation in input and output types, we select three representative tasks: code synthesis, 
code translation, and code summarization. Importantly, our imitation attacks are not limited to these tasks.
Two datasets, $D_{proxy}$ and $D_{ref}$, are used here to emulate the extracting process. 
Adversaries are only allowed to use the train split of the proxy dataset $D_{proxy}$ to construct queries. 
The reference dataset $D_{ref}$, on the other hand, is assumed to be inaccessible to adversaries.
To establish baselines for comparison, backbone models will be trained on the two datasets to form $M_{proxy}$ and $M_{ref}$, with details provided in~\S~\ref{subsec:rq1}.

\sparh{Code Synthesis (CSyn).}~CSyn in this study refers to an ``NL-PL'' task
that aims to generate specific programs based on NL descriptions. As
shown in \T~\ref{tab:task}, we use CONALA as the proxy dataset for this task,
which contains 2,879 annotations with their corresponding Python3
solutions manually collected from StackOverflow. We did not use an
online-judgment dataset such as CodeContests~\cite{CodeContests} due to the input
token length limitations. Further
discussion on this is provided in \S~\ref{sec:discussion}.

\sparh{Code Translation (CT).}~As a ``PL-PL'' task, CT involves migrating legacy
software from one language to another. As shown in \T~\ref{tab:task}, we use XLCOST as
the proxy dataset and CodeXGLUE as the reference one, where code snippets written
in Java and C\# that share the same functionality are paired together. We select 
Java as the source language and C\# as the target language.

\sparh{Code Summarization (CSum).}~As a ``PL-NL'' task, CSum generates an NL
comment that summarizes the functionality of a given PL snippet. As shown in
\T~\ref{tab:task}, we reuse the DualCODE dataset as the proxy dataset. 
We use $D_{proxy}$ and $D_{ref}$ to represent the proxy and reference dataset, respectively.

\parh{Evaluation Metrics.}~\noindent We explore two common similarity metrics for measuring the quality of
generated content. We categorize them as follows:  

\sparh{NL Content.}~For CSum, whose generated content is NL text, we follow
previous works~\cite{feng2020codebert,wang2021codet5,li2022cctest} and use the
smoothed BLEU-4 score (referred to as BLEU in the rest of this paper) to evaluate the generated NL
summarization. Given the generated text and its ground truth, BLEU examines
the number of matched subsequences, and a higher BLEU score suggests greater similarity at the token level.

\sparh{PL Content.}~The outputs of the CSyn and CT tasks are PL code snippets, which cannot be directly evaluated using NL metrics. Therefore, similar to previous
works~\cite{wang2021codet5,CodeContests,lu2021codexglue,DBLP:conf/naacl/AhmadCRC21},
we use CodeBLEU~\cite{ren2020codebleu}, a metric that takes into account token-level, structural-level, and semantic-level information.
Overall, CodeBLEU consists of four components: n-gram matching score
$BLEU$, weighted n-gram matching score $weighted\_BLEU$, syntactic AST
matching score $AST\_Score$, and semantic data flow matching score $DF\_Score$.
Specifically,
\begin{equation}
\label{equ:codebleu}
\begin{aligned}
    CodeBLEU &= \alpha * BLEU + \beta * weighted\_BLEU \\
             &+ \gamma * AST\_Score + \delta * DF\_Score
\end{aligned}
\end{equation}
\noindent where $\alpha, \beta, \gamma, \delta$ are the weights for each component.
As suggested in~\cite{wang2021codet5, lu2021codexglue}, they are all set to 0.25.
Note that both BLEU and CodeBLEU scores range from 0 to 100, with higher scores indicating a greater level of similarity.

\parh{Imitation Models.}~As noted in \S~\ref{sec:background}, adversaries can follow the classic ``pre-train and fine-tune'' paradigm,
training their imitation models via fine-tuning a well-trained backbone
model. In this work, we choose two representative models for code-related tasks: CodeBERT~\cite{feng2020codebert} and CodeT5~\cite{wang2021codet5}.
CodeT5, a variant of the text-to-text
Transformer~\cite{raffel2019exploring} model, treats all text tasks as a
sequence-to-sequence paradigm with different task-specific prefixes, making it suitable for both code understanding and code generation tasks. 
On the other hand, CodeBERT is built on a multi-layer
bidirectional Transformer encoder and pre-trained on large-scale text-code pairs
using two tasks: masked language modeling (MLM) and replaced token detection
(RTD). In the MLM task, the model predicts the original token in the masked
positions, while in the RTD task, a discriminator is used to distinguish the
replaced tokens from the normal ones.
These two models are selected for this study due to their widespread adoption and strong performance on various tasks, as evidenced by prior work~\cite{niu2023empirical,wang2022no}. Specifically, CodeBERT-base employs a 125 million parameter encoder-only architecture, whereas CodeT5-base utilizes a 220 million parameter encoder-decoder design tailored for sequence-to-sequence tasks.

\parh{Training Setup.}
To ensure the reproducibility of our results, we train the imitation model
three times for each experiment and report the median of the achieved results. In addition, we analyze 
the impact of different hyperparameters in \S~\ref{subsec:rq4}. All experiments are performed on a machine with an Intel
Xeon Patinum 8276 CPU, 256 GB of main memory, and 4 NVIDIA A100 GPUs.
By using the early stopping strategy, the training process takes an average of 5 hours and 23 minutes to complete.

\subsection{AE Generation}
\label{subsec:appr-aegener}

As discussed in \S~\ref{subsec:back-AE}, AE generation has become a common technique to improve model robustness. However, discovering AEs effectively remains challenging for advanced models. To address this, we demonstrate the feasibility of boosting AE generation for code-related tasks using an imitation model.
In contrast to the deficiency of prior methods against modern LLMs 
(detailed in \T~\ref{tab:AEmethod}), our approach using the imitation model allows for
generating AEs in an approximately white-box setting. Specifically, given the
$M_{imi}$ on hand, we rely on the attention scores to locate sensitive tokens in
an input prompt, and then apply several semantically-equivalent transformations
(derived and extended from the CCTest's codebase~\cite{cctestcode}) to iteratively mutate those
sensitive tokens until generating AEs on $M_{imi}$. These generated AEs are then
fed to the remote victim LLM to test whether they can induce failures.

To demonstrate, we apply our approach to the code summarization (CSum) task, where a successful AE
means a subtle mutation in the code input leads to a dramatically changed,
incorrect summary.
We generate potential AEs in two steps: (i) identifying the most
sensitive tokens and (ii) applying semantically-equivalent
transformation passes on those tokens. To do so, we start with an input sequence
$\mathcal{X}=[x_1,..,x_i,...,x_m]$ and generate the output sequence (code
summary) with its original attention score $Att_{ori}$. Next, we iteratively replace each
token $x_i$ with $[MASK]$ and re-generate the summary with its corresponding attention score
$Att_{i}$. We then quantify the gap score by computing $Gap_{i} = Att_{i}-Att_{ori}$
for each token. Finally, we rank all tokens according to their gap scores in descending
order to find the most sensitive tokens. 
We then check each of the top-$k$ sensitive tokens and decide if it satisfies
any of the transformation passes offered by CCTest. If it does, we apply the
transformation to generate the potential AE.\footnote{Due to page limit, we provide the full documentation
and tutorial on our website~\cite{artifact} for this step.}

\section{Research Questions and Results}
\label{sec:evaluation}

In this section, we aim to experimentally investigate the effectiveness
of extracting specialized abilities from LLMs by answering the following research questions (RQs).

\begin{itemize}[leftmargin=*,noitemsep,topsep=0pt]
    \item RQ1: How effective is the imitation attack in code-related tasks?
    \item RQ2: How do different query schemes impact the performance of the imitation attack?
    \item RQ3: How do hyperparameters affect the performance and cost of
    imitation attacks?
    \item RQ4: To what extent can the use of imitation models help generate adversarial examples against LLM APIs?
    \item RQ5: How generalizable is the imitation attack across different LLM APIs?
\end{itemize}

\subsection{RQ1: Effectiveness of the Imitation Attack}
\label{subsec:rq1}

To answer RQ1, we study the effectiveness of our launched imitation attack
on three code-related tasks: CSyn, CT, and CSum. As noted in
\S~\ref{subsec:modeltraining} and \T~\ref{tab:task}, we use the proxy
dataset to launch queries towards the target LLM and validate the performance of the imitation model
on the reference dataset.

\begin{table}[!thp]
\caption{The main results of our imitation attack. ``I/O'' stands for ``Input/Output.'' All results are presented as BLEU scores or
  CodeBLEU scores on the test split of reference datasets, where $M_{proxy}$ and $M_{ref}$
  represent the backbone models trained on the proxy and reference datasets, respectively.
  $M_{imi}$ is the imitation model trained on the collected dataset and ``API'' stands for the best original LLM result under all three query settings.
  $M_{pure}$ is the backbone model without fine-tuning. }
  \label{tab:rq1}
  \resizebox{1.00\linewidth}{!}{
  \begin{tabular}{llllllll}
  \hline

                            & I/O Type   & Model   & API                       & $M_{imi}$  & $M_{proxy}$               & $M_{ref}$  & $M_{pure}$   \\ \hline
  \multirow{2}{*}{CSyn}     & \multirow{2}{*}{NL/PL}& CodeT5  & \multirow{2}{*}{27.51} &   24.84    &  11.53            &  24.21     &  1.40     \\
                            &                    & CodeBERT&                           &   18.61    &  9.41             &  17.09     &  N/A    \\ \hline
  \multirow{2}{*}{CT}       & \multirow{2}{*}{PL/PL}& CodeT5  & \multirow{2}{*}{69.15} &   72.19    &  27.21            &  84.30     &  4.38    \\
                            &                    & CodeBERT&                           &   68.58    &  24.82            &  79.05     &  N/A    \\ \hline
  \multirow{2}{*}{CSum}     & \multirow{2}{*}{PL/NL}& CodeT5  & \multirow{2}{*}{12.90} &   17.72    &  17.25            &  18.95     &  3.84    \\
                            &                    & CodeBERT &                          &   14.09    &  12.20            &  14.87     &  N/A    \\ \hline
  \end{tabular}
  }
\end{table}

\T~\ref{tab:rq1} presents the main results of our model extraction attack.
The $M_{proxy}$ and $M_{ref}$ columns show two baseline settings in which the
backbone models are trained directly on the proxy dataset and the reference
dataset, respectively. 
The $M_{imi}$ column reports the attack accuracy, while the API column shows the LLM
performance.
It is essential to note that $M_{proxy}$, $M_{ref}$, and $M_{imi}$ have the same
model architecture (either CodeT5-base or CodeBERT-base), and
their training datasets are of equal size. We tested all three models on the
same test dataset and chose the best performance from all three query schemes.
The values in each cell represent the BLEU score for natural language texts and
the CodeBLEU score for programming language contents, as mentioned earlier in \S~\ref{subsec:modeltraining}. Furthermore,
since all three models (and the LLM APIs) are assessed using the test
split of the reference dataset, it is reasonable to observe that $M_{ref}$ (which is
trained using the train split of the same dataset) consistently achieves the
best performance across all three tasks.

From \T~\ref{tab:rq1}, it is evident that the imitation attacks are highly 
effective. First, the performance of CSyn and CT tasks is quite promising, as the
imitation models $M_{imi}$ outperform $M_{proxy}$ by an average of 57.59\% and
56.62\% on CodeT5 and CodeBert, respectively. 
Note that the proxy dataset comprises a set of input-output tuples,
whereas $M_{imi}$ is trained using the same inputs and their corresponding
outputs from the LLM API. Therefore, we attribute the superiority of $M_{imi}$ over
$M_{proxy}$ to the fact that
LLM APIs provide high-quality code snippets as outputs, which further enhance the performance of the imitation model.
Moreover, $M_{proxy}$ consistently performs the worst on all three tasks,
indicating the high value of the outputs provided by the LLMs.

\begin{figure}[!htpb]
  \centering
  \includegraphics[width=1.0\linewidth]{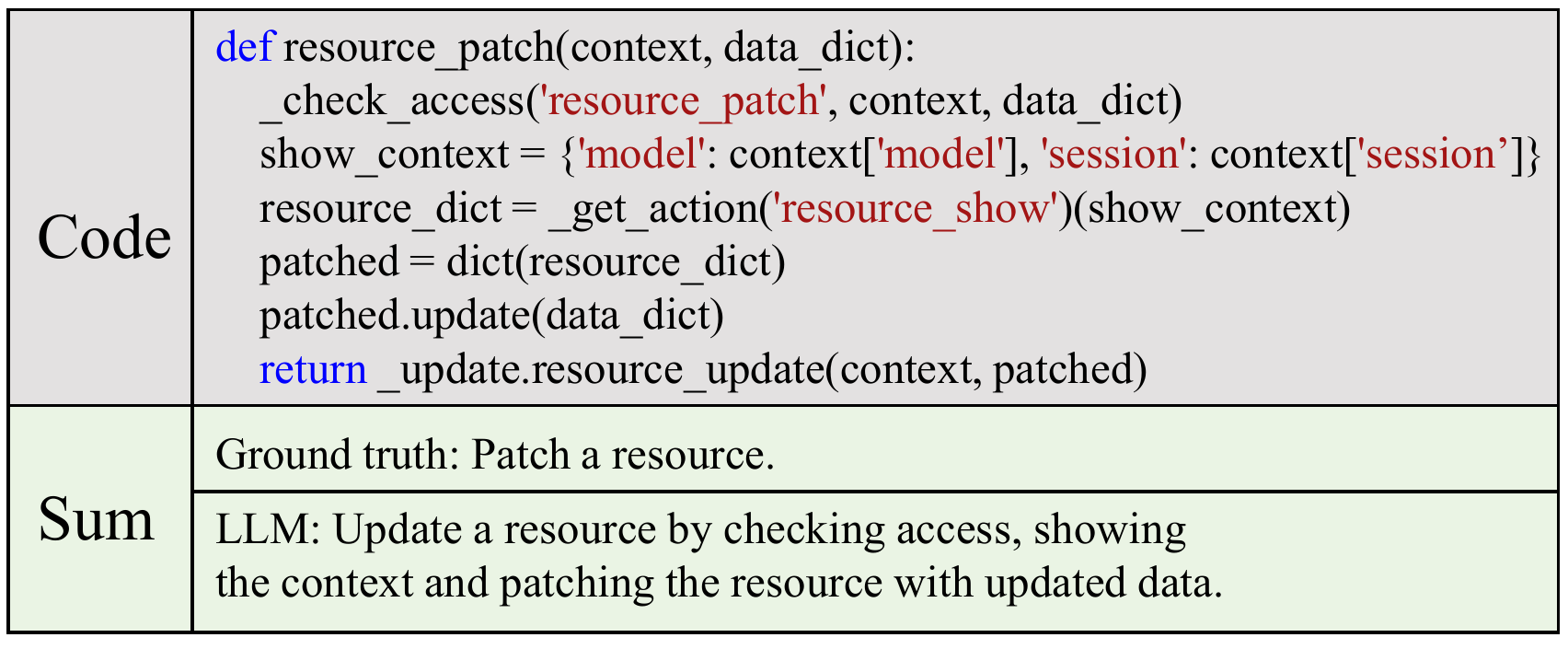}
  \vspace{-20pt}
  \caption{An example to demonstrate why LLM APIs tend to have low scores on the code summarization task, which is because the ground truth for this task uses short summaries.}
  \vspace{-5pt}
  \label{fig:codeexample}
\end{figure}

When models are trained using publicly available resources, they may perform
poorly when tested on different datasets, as evidenced by the results in the $M_{proxy}$
column of \T~\ref{tab:rq1}. This is one reason why adversaries may resort to
extracting LLMs. Encouragingly, we find that the imitation model $M_{imi}$,
when being enhanced with knowledge extracted from the LLM APIs, can largely
outperform the baseline model $M_{proxy}$ in both the CSyn and CT tasks, achieving an
average improvement of 140.3\% and 170.8\% on CodeT5 and CodeBert, respectively.
Moreover, we observe highly promising results that $M_{imi}$ can even
outperform the LLM APIs on the CT and CSum tasks, demonstrating the general
effectiveness of our imitation model.

Comparing $M_{imi}$ and $M_{proxy}$, we find that the improvement of the
imitation model $M_{imi}$ on the CSum task is less significant than the improvement on the other
tasks. However, we also notice a perplexing observation that the LLM APIs
perform even worse than $M_{proxy}$ on the CSum task.
Upon manual inspection, we discover that the APIs tend to return verbose contents
with abundant information if no additional context is provided, leading to a decrease in performance. 
For instance, in \F~\ref{fig:codeexample}, the ground truth
answer for the PL input is ``\textit{patch a resource,}'' which is concise
and straightforward. However, the answer provided by LLMs is ``\textit{Update a
resource by checking access, showing the context, and patching the resource with
updated data,}'' which is more lengthy. This phenomenon 
has been previously mentioned in~\cite{bubeck2023sparks}, where GPT-4 was utilized to
address the issues with the similarity metrics. 
However, we find this solution impractical because employing powerful LLM APIs 
such as GPT-4 as an automatic judge would still yield biased judgment results~\cite{zheng2023judging}. 
LLM judges tend to assign higher scores to lengthy responses~\cite{wang2023far}, even when conciseness suffices.

\parh{Impact on Pre-Trained Models.}~As noted in
\S~\ref{subsec:modeltraining}, both CodeT5 and CodeBERT were pre-trained on
large corpora that may overlap with our chosen test sets. To
mitigate this threat to validity, we report the baseline results of the
unmodified backbone models on the test splits in \T~\ref{tab:rq1}, in the
$M_{pure}$ column. For CodeT5, performance is substantially lower across all
three tasks compared to the other settings, indicating that the main capability
of the imitation model $M_{imi}$ is not mainly inherent from
pre-training. 
Note that CodeBERT is an encoder-only model: benchmarking its decoding
capability requires training a task-specific decoder, and its baseline
performance is thus marked as ``N/A''. 
Overall, the largely improved fine-tuning performance $M_{imi}$
indicates that pre-training alone cannot account for the models' proficiency on
three code-related tasks.

\finding{1}{Extracting specialized code abilities of LLMs through medium-sized backbone
models is effective for representative code-related tasks. The trained imitation
models achieve comparable, if not better performance than the original LLMs in
those specialized code abilities.}

\begin{table}[!thp]
\centering
\caption{Attack effectiveness using different query schemes. CBLEU denotes the
CodeBLEU metric.}
\label{tab:rq2}
\vspace{-12pt}
\resizebox{0.7\linewidth}{!}{
\begin{tabular}{llllll}
\hline
Task                   &  Model & Metric   &   ZSQ    &   ICQ         &  ZS-COT     \\ \hline
\multirow{3}{*}{CSyn}  & CodeT5 & CBLEU    &   23.39  & 23.67         & 24.84  \\
                       &CodeBERT& CBLEU    &   15.99  & 16.59         & 18.61  \\ 
                       &        & $r_{f}$\%&   2.48   & 2.40          & 4.62      \\  \hline  
\multirow{3}{*}{CT}    & CodeT5 & CBLEU    &   36.31  & 72.19         & 34.64 \\
                       &CodeBERT& CBLEU    &   37.13  & 68.58         & 34.68    \\ 
                       &        & $r_{f}$\%&   28.61  & 20.72         & 27.02     \\  \hline
\multirow{3}{*}{CSum}  & CodeT5 & BLEU     &   10.95  & 17.72         & 12.25  \\
                       &CodeBERT& BLEU     &   9.80   & 14.09         & 11.51       \\ 
                       &        &$r_{f}$\% &   0.00   & 0.15          & 0.06      \\  \hline               
\end{tabular}
}
\vspace{-15pt}
\end{table}

\subsection{RQ2: Influence of Different Query Schemes}
\label{subsec:rq2}

To answer RQ2, we aim to explore the impact of different query schemes on the
data quality and the performance of imitation attacks. Recall in
\S~\ref{sec:background}, we have introduced three query schemes. Here, we
present the comparison results on three tasks in \T~\ref{tab:rq2}. For each
setting, we represent the performance scores and the failure rates.

\parh{Imitation Performance.}~From the result listed, we can find that the
ICQ achieves better performance than the other
schemes on both CSum and CT tasks. In particular, the BLEU score on CSum
for ICQ achieves 62.83\% and 36.91\% improvement, compared with ZSQ and ZS-COT, respectively. 
These findings suggest that incorporating context information can greatly
enhance the quality of imitation training by providing more accurate answers. We
note that this finding is consistent with previous research~\cite{brown2020language,lampinen2022can} that highlights the importance
of context in natural language processing tasks. 

Moreover, ZS-COT achieves a CodeBLEU score of 24.84 on the CSyn
task for CodeT5, which outperforms ICQ by 4.94\%. We further compare four subscores
in CodeBLEU (see Equ.~\ref{equ:codebleu}) and find that the semantic data flow
matching score $DF\_Score$ of ZS-COT is 40.65, 24.12\% higher than that of the 
ICQ scheme.  
To explore the root cause of our findings, two authors of this paper manually
checked 150 ZS-COT responses for each task. Surprisingly, only 7 (4.6\%)
responses provide a meaningful rationale on CSum.\footnote{The Cohen's Kappa is
0.96, indicating that the inspection results among two authors are highly
consistent.} For the remaining responses, the rationale $r_i$ was the same as
the final answer $a_i$, indicating that ZS-COT spends a large amount of
resources but gains a limited amount of useful information. 

On the other hand, 13 (8.6\%) rationales are meaningful on CSyn, which is nearly
twice as many as those in other schemes. We attribute this to the fact that the
PL inputs in CSum and CT lack clear multi-step thinking in their questions, unlike
the NL inputs in reasoning tasks. We find that reasoning problems present their thought
processes more explicitly, while implicit thoughts in code tasks are harder for
ZS-COT to capture. Therefore, it is understandable why ZS-COT performs better on
CSyn, as its context is more closely aligned with natural language expressions
found in reasoning tasks.

\parh{Failure Rates.}~To give a comprehensive understanding of the influence of
query schemes, we also explore the difference between generated contents. As we
have described in \S~\ref{subsec:response}, we count the number of failures
among the texts and code snippets and return the failure rate $r_f\%$. For the
CSum task, all schemes share a low failure rate (less than 0.2\%), with CoT
having a slightly higher failure rate as it may contain the extra reasoning
steps in the final content that exceeds the upper bound; it may also generate a
too concise summary that is below the lower bound. 
Additionally, in comparison with ZSQ and ICQ, ZS-COT on CSyn has a notably high
failure rate of 4.62\%. This implies that the Chain-of-Thought may be less
appropriate for tasks that aim to generate code snippets. Note that all three schemes
manifest high failure rates (25\%) on CT, which is due to the nature of the CT
task. As described in \S~\ref{sec:approach}, we convert the Java programs into
C\# programs at the function-level, which diminishes some key information. To
empirically confirm this, we repeat the same grammar check experiment on the test split of the reference datasets,
which turns out to have a comparably high failure rate of
17\%.

\parh{Influence of Response Check.}~As described in \S~\ref{sec:approach}, we
design a response check algorithm to remove low-quality responses. At
this step, we explore its influence on the CSum task. We report that when
enabling this algorithm, our approach can lower the training time by
approximately 34.81\% and 14.98\% for CodeT5 and CodeBERT, respectively.
Additionally, the imitation models' performance is decreased by a maximum of
only 4\%.

\finding{2}{Query schemes have a major impact on the performance of imitation
attacks. It is vital to construct query schemes with a suitable template design.
Queries with sufficient context generally enhance the attack toward all tasks,
while the CoT only shows a noticeable boost in code synthesis.}

\subsection{RQ3: Impact of Hyperparameters}
\label{subsec:rq3}

To answer RQ3, we need to study the impact of hyperparameters. To do so, we
explore the following three aspects: 1) victim model parameters, 2) the number
of in-context examples, and 3) the number of issued queries.

\parh{Victim Model Parameters.}~According to OpenAI~\cite{davinci},
text-davinci-003 mainly provides two parameters that control its generated
outputs: $temperature$, which represents the propensity for choosing the
unlikely tokens; and $top_p$, which controls the sampling for the set of
possible tokens at each step of generation. To determine the appropriate value
for these parameters, we prompt text-davinci-003 with all 25 combinations ($5$ 
$temperature \times 5$ $top_p$) at step increments of 0.25. We use the CSum task at
this step and query those 25 sets of combinations, and the responses are then
compared with the ground truth answers to compute a BLEU score.

\begin{figure}[!htpb]
  \centering
  \includegraphics[width=1.0\linewidth]{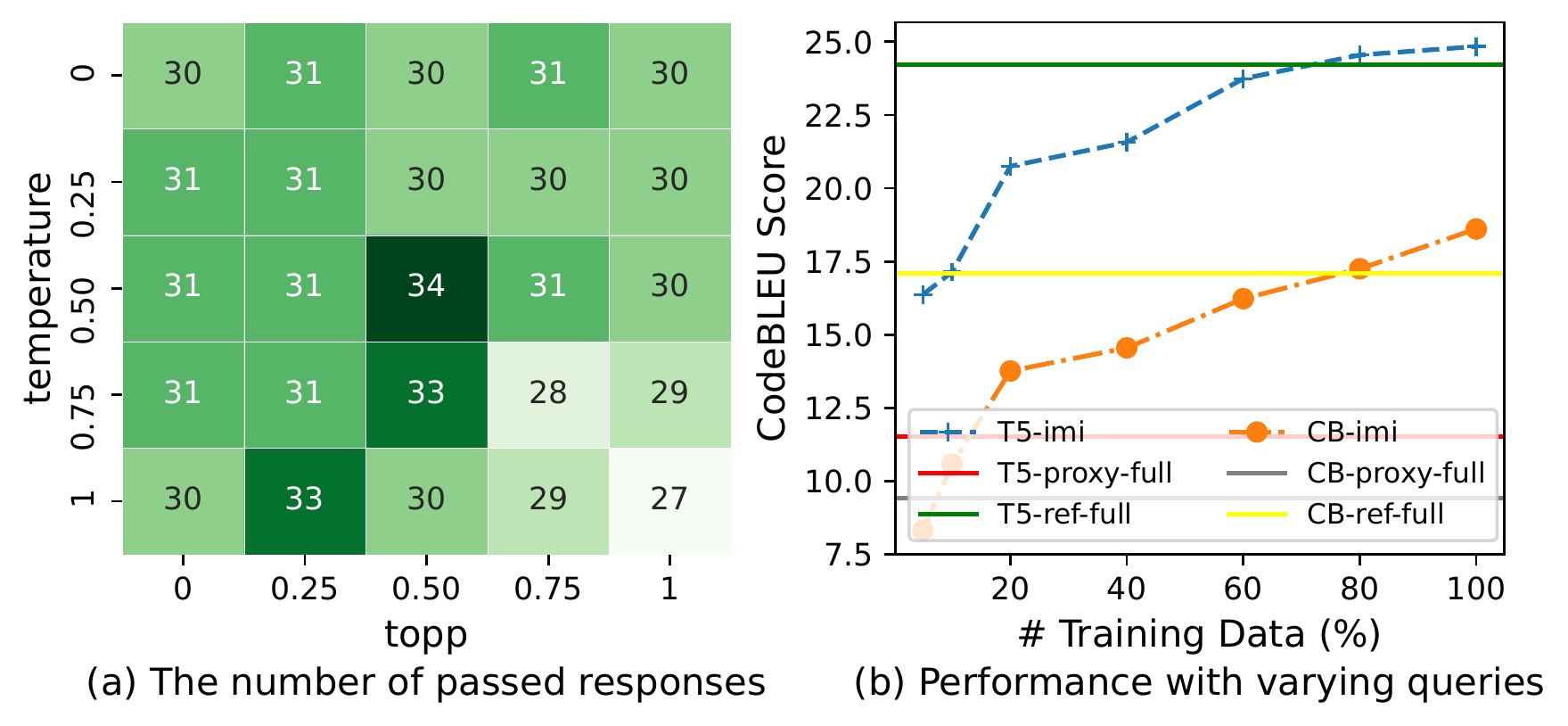}
  \vspace{-20pt}
  \caption{The impact of hyperparameters (RQ3).}
  \vspace{-5pt}
  \label{fig:passcheckbothquery}
\end{figure}

\F~\ref{fig:passcheckbothquery} (a) reports the evaluation results. For each
setting, we issue 50 queries, collect the answers yielded by the LLM, and
compare them with the ground truth (for CSum, ground truth is natural language
snippets for code summarization). An answer deems passing the check if its
similarity score is higher than the average value and otherwise as a failure. We
count the number of passing answers and present them in each cell. The
performance clearly degrades with edge values for both hyperparameters. Across
the experiments, the best parameters are seen as $top_p$ = 0.5 and $temperature$
= 0.5.

\parh{In-context Examples Number.}~As noted in \S~\ref{subsec:inc-intro}, the
number of in-context examples $E$ plays an important role in understanding the
task, and previous works~\cite{liu2021makes,chen2022improving} have used a
range of $E$ varying from two to dozens. However, this is not practical in 
code-related tasks, since the average token length of code snippets is much
longer than the natural language texts. Moreover, the victim LLM used in this
paper has a maximum token length limit (4,097) for each query, including both
inputs and outputs. Therefore, following~\cite{lampinen2022can}, we only use
example numbers ranging from 1 to 5 on the CSum task to avoid exceeding the
request limitation.

\begin{table}[!thp]
\caption{Attack performance under different number of in-context examples. T5 and CB
stand for CodeT5 and CodeBERT, respectively.}
  \label{tab:incnumber}
  \vspace{-5pt}
  \centering
\resizebox{0.8\linewidth}{!}{
  \begin{tabular}{lllllll}
    \hline
    \# In-context examples       &          & 1     & 2 & 3 & 4 & 5 \\  \hline
    \multirow{2}{*}{Model} & T5       & 14.20 & 16.51  & 17.05  & 17.72  & 16.96   \\
                           & CB       & 10.87  & 12.90  &  13.95 & 14.09  & 13.88  \\  \hline
    Cost                   &          & 11.24      &20.53   &30.97   & 42.85  &  53.29 \\  \hline
    \end{tabular}
}
\vspace{-5pt}
  \end{table}

The outcomes are shown in \T~\ref{tab:incnumber}, indicating that the
performance level escalates with an increase in \#in-context examples and
reaches its peak at $E=4$. However, there exists a discrepancy between the
improvement in performance and the accompanying cost. Although performance
gradually enhances at a dropping rate over time, sustaining optimal performance
incurs linearly growing costs. This is because every new example added demands
extra tokens associated with that example for all queries, even those that have
already accomplished satisfactory performance. Considering this trade-off,
adversaries must carefully assess their options. For imitation, we suggest
fixing the example number to three. We regard this as a reasonable trade-off that
adversaries are willing to embrace during implementation.

\parh{Number of Queries.}~Considering that the performance of imitation attacks
heavily relies on the number of collected
responses~\cite{orekondy2019knockoff,wallace2020imitation}, an evident trade-off
exists between the number of queries and the performance of the imitation model.
Too few queries can lead to poor performance while too many queries may cause an
unaffordable cost, and neither scenario is desirable for adversaries. We now study 
the impact of \#queries on imitation attack performance.

The evaluation results are shown in \F~\ref{fig:passcheckbothquery}(b),
where T5 and CB represent CodeT5 and CodeBERT, respectively. Due to the
limited space, we report the evaluation results on the CSyn task. Other
tasks share similar findings; see~\cite{artifact} for the full results.
In short, when the adversary-collected dataset is equivalent in size to that
of the model owners' training dataset, the imitation model surpasses the model
trained on the proxy dataset (``T5/CB-proxy-full'') and attains comparable
performance to the model trained on the reference dataset (``T5/CB-ref-full'').
This clearly shows the value of conducting imitation attacks, as the adversaries
in real-world usually face the problem of lacking appropriate training data.
We note that for the CSyn task, the increasing number of queries could help
address the insufficient data problem and boost performance. With only 5\%
queries, our imitation model $M_{imi}$ trained on the collected data can achieve
a notable improvement of 41.97\% on CodeT5 as compared to $M_{proxy}$, and the
performance keeps increasing and reaches a peak with a 115.43\% improvement.

Upon analyzing the collected dataset, we view that the internal diversity and
high quality of gathered data were the reasons behind the improvement. For
example, when given the same NL instruction ``\textit{decode a hex
string `4a4b4c' to `UTF-8'}'', the model trained on the proxy dataset returns
the ``\textit{hex = hex_decode(hex) return hex}'' that is full of repeated
tokens ``\textit{hex}''. In contrast, $M_{imi}$ generates
``\textit{bytes.fromhex(`4a4b4c').decode(`utf-8')}'', a compilable code snippet
that uses the built-in APIs correctly.
One remaining question is whether such improvement can keep increasing to a
considerably large extent. Unfortunately, we notice that the boost from the
increasing number of queries gradually diminishes with its growing size, which
reflects the margin effect of \#queries.

\finding{3}{The output/sampling hyperparameters have an observable (yet not
significant) impact on the attack performance. In contrast, \#queries and
\#in-context examples notably affect the attack performance.}

\subsection{RQ4: Boosting AE Generation}
\label{subsec:rq4}

After attackers obtain a well-performing imitation model $M_{imi}$, this
RQ investigates the feasibility of generating AEs to further
exploit the target LLM and manipulate its outputs. 

In this RQ, we study the code summarization task (CSum), where a successful AE
for this task constitutes a subtle perturbation to the input code 
which results in a significantly altered, incorrect summary from the model.
We note that the same method can be easily extended to other
tasks. As discussed in \S~\ref{subsec:rq1}, the natural language descriptions in
the CodeSearchNet dataset are highly ambiguous, which increases the difficulty
in distinguishing the real AEs. Therefore, we use the Leetcode~\cite{leetcode}
dataset as the test dataset, as it provides clearer explanations for the given
code snippets.

\begin{figure}[!htpb]
  \centering
  \includegraphics[width=1.01\linewidth]{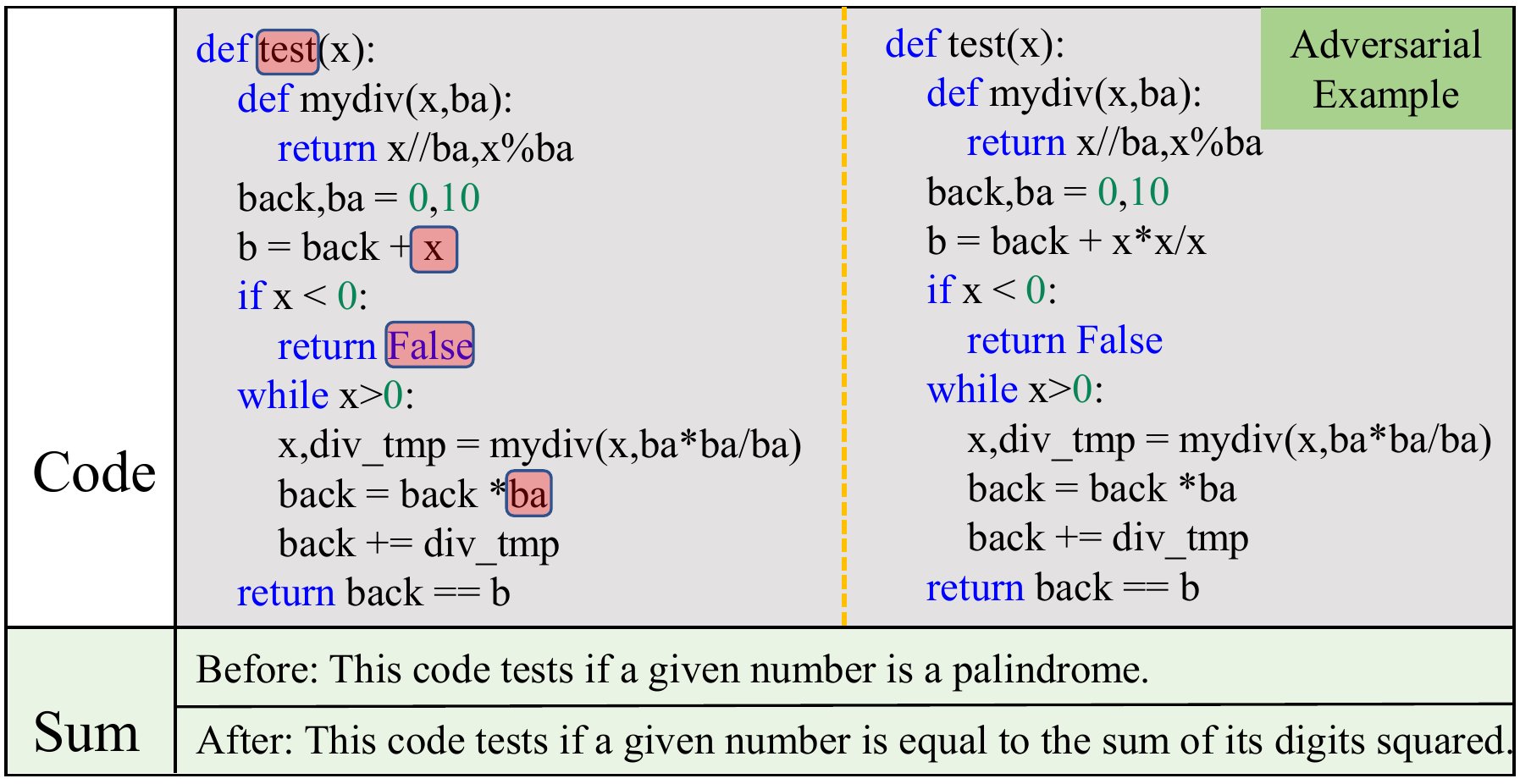}
  \vspace{-15pt}
  \caption{Adversarial examples generation.}
  \vspace{-10pt}
  \label{fig:AEexample}
\end{figure}

As introduced in \S~\ref{subsec:appr-aegener}, there are two primary steps in generating
the potential adversarial examples: (i) Finding the most vulnerable tokens. 
(ii) Applying the semantically-equivalent transformation passes on selected tokens.
We illustrate the process in \F~\ref{fig:AEexample}, where the original input code aims to
test whether a given number is a palindrome. The selected tokens are ``test, x, False, ba'' (in red text box). Since variable ``x'' satisfies the mathematical
constant transformation in CCTest, we replace it with ``x*x/x'', and it turns
out to be an AE for the text-davinci-003 API. 
It is important to note that AE generation for NLP tasks typically takes hours to produce a single instance~\cite{zou2023universal}. In contrast, our proposed method can generate adversarial examples in just seconds to minutes, depending on the input length and the choice of the imitation backbone model. This efficiency arises from the use of an imitation model to guide the adversarial example generation. As the imitation model is much smaller than the target language models, inference is considerably faster.

Besides, we also compare our approach with several previous works introduced in \S~\ref{subsec:back-AE} to demonstrate its
effectiveness.
Overall, these works are considered state-of-the-art (SOTA) AE generation methods for code models under
white-box or black-box settings, and we follow their original settings to
conduct the experiments below.

\begin{table}[!thp]
  \caption{Comparison for different adversarial attacking methods. Sem EQ, 
  SAE rate, and UAE rate stand for semantically equal, stable AE rate and unstable
  AE rate.}
    \label{tab:AEmethod}
    \vspace{-10pt}
  \setlength{\tabcolsep}{2pt}
    \centering
  \resizebox{0.9\linewidth}{!}{
    \begin{tabular}{lllll}\hline
    
      Method                            & Type       & Sem EQ?   & SAE rate & UAE rate \\ \hline
      CodeAttack                        & Whitebox   & False     & 1.11 \%     &  4.44 \%    \\
      Radar                             & Blackbox   & True      & 0    \%     &  1.47 \%  \\ 
      CCTest                            & Blackbox   & True      & 0    \%     &  1.13 \%  \\
      Ours                              & $M_{imi}$-enabled Whitebox   & True      & 9.5  \%     &  4.78\%   \\ \hline

    \end{tabular}
  }
  \vspace{-5pt}
\end{table}

We generate up to 90 potential AEs for each of the prior works and our method,
and use them to query text-davinci-003. Then, we manually inspect the responses
and report the results in \T~\ref{tab:AEmethod}. Note that OpenAI models are
non-deterministic~\cite{davinci}, meaning that identical inputs can yield
different outputs. Setting $top_p = 1$ and $temperature = 0$ will make the
outputs mostly deterministic, but a small amount of variability may remain.
Therefore, we repeat querying each AE three times and divide them into two
categories: stable AE (SAE) and unstable AE (UAE). The former demonstrates that
this example can be triggered every time we query (more desirable), while the
latter indicates that the attack succeeds for at least one time. Both
types of AEs have practical implications, as real-world users may query a model
multiple times to gain confidence in the result, or just once without repeating.
It is thus evident that the former users are presumably
vulnerable under UAE and constantly vulnerable under SAE, whereas the latter is constantly vulnerable under SAE.
It is worth noting that the example shown in \F~\ref{fig:AEexample} represents a stable AE, meaning it can be triggered each time we query the text-davinci-003 API.

\T~\ref{tab:AEmethod} reports highly encouraging results, indicating that our
method is effective in finding potential AEs in such a challenging setting.
Among methods that preserve input semantics (a generally more desirable
property), CCTest fails to trigger any incorrect content, while Radar only
generates one AE which cannot be triggered stably. In contrast, our method can
find both stable and unstable AEs. We deem that the imitation model
$M_{imi}$ provides valuable guidance to the process of AE generation.

Additionally, we notice that CodeAttack, as a SOTA \textit{white-box} attack method, can also trigger some misleading content.
Specifically, it discovers stable and unstable adversarial examples at rates of 1\% and 4\%, respectively. 
However, we find
that it does not preserve the semantic equivalence of the input, as it would
aggressively replace or delete sensitive tokens. This renders nearly one-third
of the code snippets unable to be compiled without any error~\cite{jha2022codeattack}. Furthermore,
it is evident that a white-box attacking method is not applicable to real-world
LLMs.

\finding{4}{The imitation model provides useful information, e.g., attention score, to facilitate the generation of adversarial examples against LLMs. 
With this information, we successfully discover multiple AEs that can be stably triggered on LLM APIs.}

\subsection{RQ5: Generalizability}
\label{subsec:rq5}

In this RQ, we investigate the generalizability of our imitation attack. To do
so, we follow an identical process to query other LLM APIs and train our imitation model on the collected dataset.
As we introduced in \S~\ref{sec:approach}, we conducted all experiments on
text-davinci-003 due to its great capacity. OpenAI also provides gpt-3.5-turbo,
which is highly optimized for chat.
This model is also popular, due to its relatively lower cost (1/10th of the
cost) in comparison to the davinci series of models. 
Notably, we exclude GPT-4 here since its API remains publicly
unavailable, and its web service has a strict rate limit.
While it is unclear about
the internals of gpt-3.5-turbo, we believe evaluating two highly popular and
representative LLMs (text-davinci-003 and gpt-3.5-turbo) is sufficient to
demonstrate the generalizability of our approach.

\begin{table}[!thp]
  \caption{Comparison for additional LLM APIs. TD-003 and GPT-35 stand for ``text-davinci-003'' and ``gpt-3.5-turbo''.}
    \label{tab:otherAPI}
    \vspace{-12pt}
    \centering
  \resizebox{0.68\linewidth}{!}{
    \begin{tabular}{l|ll|ll} \hline
      & \multicolumn{2}{c}{API}        & \multicolumn{2}{c}{IMI} \\ \hline
      & TD-003      & GPT-35             & TD-003      & GPT-35      \\
 CSyn & 27.51       & 24.11      & 24.84      & 22.85      \\
 CT   & 69.15       & 65.33      & 72.19      & 67.40      \\
 CSum & 12.90       & 12.2       & 17.72      & 16.51      \\ \hline
 \end{tabular}
  }
  \vspace{-7pt}
\end{table}

\T~\ref{tab:otherAPI} presents consistently promising results when attacking
gpt-3.5-turbo. Similar to RQ1, we report the best result among all
query strategies ($M_{imi}$ performance is in the ``IMI'' column). The values
in each cell represent the BLEU score for NL texts and the
CodeBLEU score for PL contents.
Notably, gpt-3.5-turbo achieves competitive performance on all three tasks compared
to text-davinci-003 (in the ``API'' column), with an average decrease of
7.78\%. Overall, we find that the imitation models trained on data collected
from gpt-3.5-turbo achieve approximately 92.84\% of the performance of those trained
on text-davinci-003, demonstrating the high generalizability of our imitation
attack method. As an implication, attackers in reality may prefer to launch
imitation attacks toward gpt-3.5-turbo, which is more cost-effective and offers
comparable imitation model performance.

\finding{5}{Our imitation attack method exhibits encouraging generalizability
and can adapt to different LLM APIs without extra adaptiveness. This
illustrates potentially more severe threats in reality, as attackers in reality
may smoothly transfer the attack to other LLM APIs with lower cost.}

\section{Discussion}
\label{sec:discussion}

\parh{Threats to Validity.}~The findings of this work face certain threats to validity that merit acknowledgment. First, the generalizability of the results may be limited given the specific LLMs and backbone models evaluated. While text-davinci-003 and gpt-3.5-turbo are selected as representative LLMs, more advanced models such as GPT-4~\cite{gpt4} and Claude2~\cite{claude2} are not assessed. Additionally, we consider only two backbone models (CodeT5 and CodeBERT); recently proposed alternatives such as Starcoder~\cite{StarCoder} and WizardCoder~\cite{WizardCoder} are not tested. It remains unclear how effective the proposed imitation attacks would be against these alternative LLMs and backbones. Second, the performance measurements obtained may depend heavily on the specific tasks selected for analysis. While we choose three representative code tasks, outcomes could vary substantially on different tasks such as code completion.
Overall, more analysis on diverse models and tasks would strengthen conclusions.

\parh{Dataset Choice.}~As mentioned in \S~\ref{sec:approach}, we do not choose
those common online-judgement datasets such as CodeContests~\cite{CodeContests}
for imitation attacks. This is because the average length of its NL description
is over hundreds of tokens (not including the corresponding explanation). Our
preliminary study shows that this is often too lengthy for medium-sized models
to learn. Additionally, recall that we explored the in-context query scheme for
each task, which requires the length to be several times larger than the
original question. Hence, lengthy queries would exceed the max token length
limit frequently, thereby failing the query process.

\parh{Semantically Equal Measurement.}~In this research, we use the CodeBLEU
score to assess the caliber of the produced code snippets. However, these
methods are imperfect since they are built on the token level and the AST level
rather than the semantic level, and this leads to the scenario that two
semantically equal code snippets may have a relatively low CodeBLEU score. Some
methods like dynamic testing~\cite{grigorenko1998dynamic} or symbolic
execution~\cite{poeplau2020symbolic,chen2022symsan} have been proposed for
semantics-level comparison; however, such methods are either too heavy to be
practical or even impractical for code snippet which does not have a clear input
and output. 
In recent studies, pass@k has been proposed as a metric for evaluating the quality of generated code~\cite{chen2021evaluating}. This metric measures the percentage of generated code snippets that pass all given test cases. However, the strict requirements of the test cases pose a challenge to its application in our research context.
In sum, we clarify that it is common to use CodeBLEU scores as a metric for assessing the quality of generated 
code and to demonstrate that the model is capable of producing better code~\cite{wang2021codet5,CodeContests,lu2021codexglue,DBLP:conf/naacl/AhmadCRC21},
and we leave exploring other metrics for future work.

\parh{Mitigation with Watermarking.}~Watermarking is a common technique for 
protecting intellectual property (IP). In the context of neural networks,
watermarking methods may involve subtly modifying the text/code (e.g., distributions of
certain synonyms) to embed IP
information~\cite{he2022cater,DBLP:conf/aaai/HeXLWW22,li2023protecting}. Model owners can then
verify their IP using statistical tests, such as Student's t-test. 
LLM vendors may have embedded working prototypes of watermarks in their
APIs~\cite{OpenAIWM}, though the details are unclear. However, we envision that
attackers may consider its potential enforcement, and pursuing attacks
imprudently would be improper, especially when imitation models are later used
for commercial purposes. For instance, Google has been accused of training its
AI chatbot Bard on data from OpenAI's ChatGPT without
authorization~\cite{gsteal}. Therefore, it is possible that extracting LLM's
specialized code abilities may not be as concerning as our paper suggests. Looking
ahead, we advocate better use of watermark and other relevant techniques to
mitigate model extraction attacks.

\parh{Reflection of Our Findings.}~The present study demonstrates the feasibility of extracting specialized code abilities of LLMs through imitation attacks. 
By employing various query schemes, we show that the imitation models can achieve comparable performance to the original LLMs in three code tasks and can also enhance downstream applications such as adversarial example generation. 
Our findings hold significant implications for the research community. 
First, the ability to extract LLMs' specialized code abilities poses a serious threat to the LLM vendors who must safeguard their intellectual property. 
Second, the effectiveness of the imitation attacks is heavily influenced by the query schemes employed, highlighting the need for a balanced approach that optimizes both cost and performance. 
Third, the imitation models can positively impact downstream applications such as adversarial example generation, which can further enhance the robustness of LLMs. Given the significance of these findings, further research is warranted in this area, and we hope that our study will inspire future investigations.

\section{Conclusion}
\label{sec:conclusion}

In this paper, we experimentally investigated the effectiveness of extracting specialized code abilities from LLMs using common medium-sized models.
To do that, we designed an imitation attack framework that comprises query generation, response check, imitation training, and downstream malicious applications.
as well as provide useful information to facilitate the generation of adversarial examples against LLMs.
We summarized our findings and insights to help researchers better understand the threats posed by imitation attacks.

\bibliographystyle{ACM-Reference-Format}
\bibliography{sample-base,bib/code,bib/cot,bib/imitation,bib/symbolic,bib/zj}

\end{document}